\documentclass[preprint,12pt]{elsarticle}
\usepackage[utf8]{inputenc}
\DeclareUnicodeCharacter{00B3}{\textsuperscript{3}}

\usepackage{amssymb}
\usepackage{amsmath}

\usepackage{xcolor}
\usepackage[normalem]{ulem}
\usepackage{soul}

\journal{Physics of the Dark Universe}

\begin{document}

\begin{frontmatter}



\title{Fundamental oscillations as a tool to distinguish boson stars from neutron stars and black holes}

\author[label1]{Swarnim Shirke\corref{cor1}}
\ead{swarnim@iucaa.in}
\cortext[cor1]{Corresponding Author}
\affiliation[label1]{organization={Inter-University Centre for Astronomy and Astrophysics},
            addressline={Ganeshkhind, Pune University Campus},
            city={Pune},
            postcode={411007},
            state={Maharashtra},
            country={India}}
\author[label1,label2]{Bikram Keshari Pradhan} 
\affiliation[label2]{organization={IP2I Lyon, University Claude Bernard Lyon 1, CNRS/IN2P3},
            addressline={UMR 5822},
            city={Villeurbanne},
            postcode={69622},
            state={},
            country={France}}
\author[label1]{Debarati Chatterjee} 
\author[label3]{Laura Sagunski} 
\author[label3]{J\"urgen Schaffner-Bielich} 

\affiliation[label3]{
            organization={Institute for Theoretical Physics, Goethe University},
            addressline={Max von Laue Str. 1}, 
            city={Frankfurt am Main},
            postcode={60438}, 
            state={},
            country={Germany}
            }

\begin{abstract}
Massive boson stars are self-gravitating configurations of self-interacting scalar fields and can be modeled by a massive scalar field with a quartic self-interaction potential. It has been shown that the equation of state and static structure properties, such as mass and radius, follow scaling relations independent of microscopic dark matter properties. In this work, we demonstrate for the first time that non-radial fundamental ($f$-)mode characteristics also follow a scaling in the strong interaction limit, opening up the outstanding prospect of evaluating the mode properties for boson stars for arbitrary masses spanning the scalar dark matter parameter space allowed by current observations. We provide the scaling relations within full general relativity and obtain the mode characteristics corresponding to the maximum boson star mass configuration. We apply these to determine the $f$-mode properties for boson stars solely as a function of their mass and compactness, which allows distinguishing them from those of neutron stars and black hole quasinormal modes in comparable mass range. In particular, we show that the frequencies are always lower than those of corresponding black holes of the same mass by a factor of 4.5. This provides a smoking gun for the distinguishability of boson stars from other compact objects using gravitational wave observations. 
\end{abstract}




\end{frontmatter}


\section{Introduction} \label{sec:intro}

Dark matter (DM) and its nature are among the most intriguing open questions of physics. DM forms $\sim 25\%$ of the energy budget of our universe. However, its particle nature has remained elusive. DM candidate masses span 90 orders of magnitude from ultralight fuzzy DM to massive compact halo objects (MaCHOs) and primordial black holes (PBHs)~\cite{Bertone2018}. Historically, the most popular candidate for DM has been Weakly Interacting Massive Particles (WIMPs). However, years of fruitless experimental search for WIMPs, discovery of Higgs Boson, prediction of axions to solve the strong CP violation, failure of cold DM paradigm on smaller scales, etc. has renewed interest in scalar bosonic particles like axions, axion-like particles (ALPs), fuzzy DM, etc. as candidates for DM (see~\cite{Hui2017} for review on bosonic DM).

Boson stars (BSs), self-gravitating stellar configurations of bosonic particles (see~\cite{Visinelli2021, LieblingPalenzuela2023} for an updated review on BSs), are a consequence of the existence of such scalar fields in the Universe. These can form from primordial scalar fluctuations, gravothermal collapse in an early matter-dominated era~\cite{Ralegankar2024}, or via a process called gravitational cooling~\cite{Seidel1994}. They can also be formed via trapping of DM in objects like stars and planets~\cite{ ellis2018DMaccumulation, Horowitz2019ns, Horowitz2020sun, Horowitz2020earth}. BS is one of the possible exotic compact objects (ECOs) that can mimic other compact objects like black holes (BHs) and neutron stars (NSs) and even source PBHs formation~\cite{Khlopov1985}. BSs, if they exist, would also contribute to the total DM budget (see~\cite{CardosoPani2019} for a review on ECOs) of the universe and their existence would be a key to the DM problem, revolutionizing our understanding of the universe.

Conventional electromagnetic telescopes can probe BSs only if they are accreting matter~\cite{Olivares2020, Rosa2022, Rosa2023}, keeping non-accreting ones out of reach. However, the successful detection of gravitational waves~\cite{GWDetection2016} has opened up a new window to probe the dark sectors of the universe. Further, the planning of the next-generation detectors covering a wide frequency range makes the study of all possible gravitational wave (GW) sources of BSs very crucial and timely.
There have been considerable efforts to find observational signatures that will be able to distinguish BSs from BHs. Some studies suggested the distinguishability of BSs and BHs as well as NSs based on tidal deformability~\cite{Cardoso2017, Sennett2017, McDaniel2020}, spin-induced multipole moments~\cite{Krishnendu2017, Krishnendu2019, Pacilio2020, Vaglio2023}, $I$-Love-$Q$ relations~\cite{Maselli2017} and mergers~\cite{Cipriani2024}. 

The study of quasinormal modes (QNMs) is crucial as it carries information about the nature of the underlying compact object~\cite {Bustillo2021, Capano2023}. There have been a handful of studies on BS QNMs~\cite{Kojima1991, Yoshida1994, Balakrishna1998, Macedo2013a}, however, these considered weakly interacting systems. Our focus in this work are massive BSs in the strong-interaction limit (refer to Sec.~\ref{sec:model} for details). Flores et al.~\cite{Flores2019} is the first work to explore $f$-modes, which couple the strongest to GWs, for massive BSs in the strong-interaction limit. The study was restricted to NS mass BSs, and the parameters were restricted by the bounds on self-interaction cross section $0.1 < \sigma/m < 10$ [cm$^2$/g]. Another work~\cite{Celato2025}, that appeared during the completion of this work, again was restricted to select parameters in the same regime and explored $f$-mode universal relations.

In this letter, we show for the first time that the $f$-mode eigenfrequency follows a scaling relation in the strong-interaction limit, proposing a new signature based on $f$-mode oscillations to delineate massive BSs from BHs and NSs. This scaling is significant, as it allows commenting on the $f$-mode characteristics of BSs with arbitrary self-interacting scalar DM parameters.  Using this, we make a detailed comparison of BS $l=2$ QNMs with those of BHs and NSs to comment on the distinguishability. 
Throughout this work, we use the convention $\hbar=c=1$ and $G=1/M_{Pl}^2$, where $M_{Pl} = 1.22 \times 10^{22} \text{ MeV} = 2.18 \times 10^{-8} \text{ kg}$ is the Planck mass.

\section{Model} \label{sec:model}

For pure Bosonic dark stars, we describe DM by a scalar field ($\phi$) self-interacting via quartic self-interactions as described in~\cite{Colpi1986, Flores2019}. The Lagrangian is given by
\begin{equation}
    \mathcal{L} = \frac{1}{2}\partial_{\mu}\phi^*\partial^{\mu}\phi + V(\phi)~,
\end{equation}
\begin{equation}
    V(|\phi|) = \frac{1}{2}m^2|\phi|^2 + \frac{1}{4}\lambda|\phi|^4~,
\end{equation}

where $\lambda$ is the self-interaction parameter and $m$ is the DM particle mass. The self-interaction is negligible only if $\lambda|\phi|^4/m^2|\phi|^2 \ll 1$~\citep{Colpi1986}, where $|\phi| \sim M_{Pl}$. 
A dimensionless parameter is defined as $\Lambda \equiv \lambda M_{Pl}^2/4\pi m^2$.

For $\Lambda \gg 1$, Colpi et al.~\cite{Colpi1986} showed that massive stars can be formed with mass resembling that of a degenerate fermionic star. In this case, $M_{max} = 0.22 \sqrt{\Lambda}M_{Pl}^2/m = 0.062\sqrt{\lambda}M_{pl}^3/m^2$~\cite{Colpi1986}. Moreover, $\Lambda \gg 1$, can be achieved for $\lambda \ll 1$. In this limit, even if $\lambda \ll 1$, the interaction is important, and the macroscopic properties are dictated by the strength ($\sqrt{\lambda}/m^2$).

We consider here the case of self-interacting massive BSs, for which  $\Lambda \gg 1$, which is referred to as the strong-interaction limit~\cite{Seoane2010}. In this limit, the Einstein-Klein-Gordon system of equations for the scalar field resembles that of a perfect fluid star with the effective equation of state given by~\cite{Colpi1986, Karkevandi2022, Cipriani2024},
\begin{equation}
    p = \frac{m^4}{9\lambda}\left(\sqrt{1+\frac{3\lambda}{m^4}\rho}-1\right)^2~. 
\end{equation}

\section{Scaling}\label{sec:scaling}
The mass of massive BSs is known to scale as $\sqrt{\lambda}M_{pl}^3/m^2$~\cite{Colpi1986} while its radius scales as $\sqrt{\lambda}M_{Pl}/m^2$. We define a quantity as done in~\cite{Cipriani2024}:
\begin{equation}\label{eqn:x_def}
    x \equiv \sqrt{\lambda}/m^2~.
\end{equation}

Colpi et al.~\cite{Colpi1986} showed that the energy density and pressure scale as $\rho' = \rho x^2,~ p' = px^2$
to obtain the scaled version of the equation of state 
\begin{equation}
    p' = \frac{1}{9}(\sqrt{1+3\rho'}-1)^2~.
\end{equation}
The dimensionless mass ($M'$) and radius ($R'$) can then be defined for BSs as $M' = M/(xM_{Pl}^3),~R' = R/(xM_{Pl})$ and can be used to obtain scaled TOV equations.

Here, we are interested in the $f$-mode frequencies and damping times. We introduce the scaling for the complex $f$-mode eigenfrequency ($\omega$) for the first time which is given by 
\begin{equation}\label{eqn:fmode_scaling}
    \boxed{\omega=\omega'/(xM_{Pl})}~.
\end{equation} 
This means the $f$-mode frequency ($f$) and damping time $\tau$ scale as
 \begin{align}\label{eqn:fmode_scaling_f}
     f &= f'/(xM_{Pl}),~\tau = \tau'(xM_{Pl})~.
\end{align}
We explicitly derive this scaling using the $f$-mode perturbation equations for the case of Cowling approximation, where the background metric is held constant (see~\ref{sec:appendix_fmode_cowling}) as well as in full general relativity (see~\ref{sec:appendix_fmode_gr}). This is one of the main results of this work. We now discuss the implications of this scaling.

\section{$f$-modes}\label{sec:f_modes}
When the scaling as mentioned in Eq.~\ref{eqn:fmode_scaling} is used, the $f$-mode equations become completely independent of the parameter $x$ containing the DM parameters (see appendices). As a preliminary check, we reproduce the results for $f$-mode characteristics, i.e., frequency and damping times, from Flores et al.~\cite{Flores2019}.
This work restricted their study to stellar mass BSs and used select values of $m$ and $\lambda$ to study $f$-modes.
These values are consistent with the $\sigma/m$ constraints on self-interacting DM and only restricted the work to BSs falling in the mass range of 1-6$M_{\odot}$. The lower bound on $\sigma/m$ is not robust, and one can have BSs with arbitrary mass depending on the allowed DM parameter space. We apply the scaling $f' = fxM_{Pl}$ and $\tau' =\tau/xM_{Pl}$ and plot the scaled frequency and damping time, and find that upon scaling the parameters, all the curves coincide (not shown here). Thus, there is no need to study $f$-modes separately for different values of ($\lambda,m$) in the parameter space of massive BSs in strong interaction limits.

\begin{figure}
    \centering
    \includegraphics[width=0.6\linewidth]{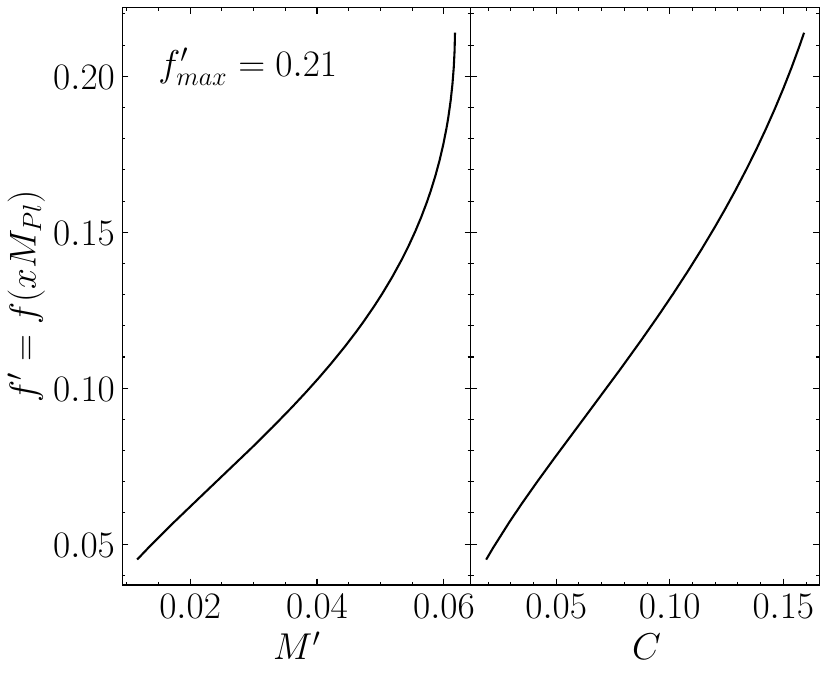}
    \caption{The scaled $f$-mode frequency ($f'=f(xM_{Pl})$) for massive BSs in the strong interaction limit ($\Lambda \gg 1$) as a function of scaled mass $M'=M/(xM_{Pl}^3)$ and compactness $C$. These are unique curves and the scaling relations can be used to obtain the corresponding $f-M$ and $f-C$ for any microscopic parameters ($\lambda$, $m$). For a given set of parameters, the $f$-mode frequency corresponding to the maximum compactness configuration is the highest and is given by $f_{max}=f'_{max}/(xM_{Pl}) = 0.21/(xM_{Pl})$.}
    \label{fig:f_m_c}
\end{figure}

In Fig.~\ref{fig:f_m_c}, we first plot dimensionless $f'$ as a function of $M'$ and compactness $C=M/R=M'/R'$. We get unique curves from primed quantities. This is because the $f$-mode equations and perturbation equations are completely independent of model parameters when written in scaled coordinates. We can then use scaling $f'=f(xM_{Pl})$, $M'=M/(xM_{Pl}^3)$ and $C'=C$ to obtain the curves for any set of $(\lambda, m)$ parameters. 

$f'$ is an increasing function of $M'$ and $C$ and reaches a maximum value of 
 \begin{equation}
     \boxed{f'_{max}=0.21}
 \end{equation} 
for the maximum mass configuration of BS. Thus, for arbitrary DM parameters, the maximum possible BS $f$-mode frequency corresponding to the maximum compactness configuration is given by $f_{max} = 0.21m^2/(\sqrt{\lambda} M_{Pl})$ (see Eq.~\ref{eqn:fmode_scaling_f}). The increasing behavior is expected as the frequency is known to increase with the average density ($f \sim \sqrt{M/R^3}$), and in the case of BSs, it is known that mass increases while the radius decreases as we increase the central density. 

\begin{figure}
    \centering
    \includegraphics[width=0.6\linewidth]{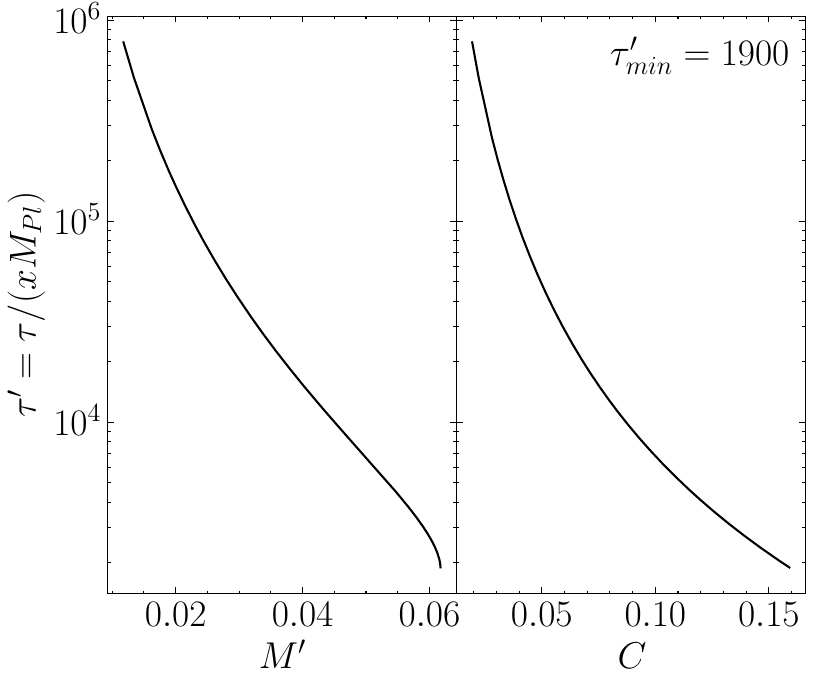}
    \caption{The scaled $f$-mode damping time ($\tau'=\tau/(xM_{Pl})$) for massive BSs in the strong interaction limit ($\Lambda \gg 1$) as a function of scaled mass $M'=M/(xM_{Pl}^3)$ and compactness $C$. These are unique curves and the scaling relations can be used to obtain the corresponding $\tau-M$ and $\tau-C$ for any microscopic parameters ($\lambda$, $m$). For a given set of parameters, the $f$-mode damping time corresponding to the maximum compactness configuration is the lowest and is given by $\tau_{min}=\tau'_{min}(xM_{Pl}) = 1900(xM_{Pl})$.}
    \label{fig:tau_m_c}
\end{figure}
 
Next, we show $\tau'$ as a function of $M'$ and $C'$ in Fig.~\ref{fig:tau_m_c}. $\tau'$ is decreasing function of $M'$ and $C'$. We use the log scale for $\tau'$ as it varies by orders of magnitude. Again, we find that scaled relations are independent of model parameters, i.e., for a given $M'$ (or $C'$), we have a unique $\tau'$. The minimum value of $\tau'$ corresponding to the maximum mass configuration is 
 \begin{equation}
     \boxed{\tau'_{min} =1900}~.
 \end{equation}
 This single solution is sufficient to reproduce all the results and for the entire parameter space in the strong coupling regime ($\Lambda \gg 1$) using the relations $\tau'=\tau/(xM_{Pl})$, $M'=M/(xM_{Pl}^3)$ and $C'=C$. For a given set of parameters, the minimum possible BS $f$-mode damping time corresponding to the maximum compactness configuration is given by $\tau_{min}= 1900\sqrt{\lambda}M_{Pl}/m^2$ (see Eq.~\ref{eqn:fmode_scaling_f}).

Combining these two results, we find that $N=f\tau = f'\tau'$ is a decreasing function of mass and compactness. This is expected as $\tau$ falls on a logarithmic scale. $N$ represents the number of oscillations that the BS undergoes before the amplitude of the oscillations damps by a factor of $e$. Thus, corresponding to the maximum compactness configuration, we get a minimum of
 \begin{equation}
     \boxed{N_{min} = 406}
 \end{equation}
 oscillations. Thus, BSs with any microscopic model parameters and compactness would undergo at least $N_{min}$ number of oscillations if $f$-modes are excited.

 \section{Comparison with neutron stars and black holes}\label{sec:comparison}
It would be interesting to compare the BS frequencies with those of the QNMs of other astrophysical compact objects. Here, we compare with the fundamental frequencies of NSs and BHs, which are dominant modes in the context of GW. 

For the case of BSs, the $f$-mode frequency as a function of mass can be derived using the scaling relations. We have $f = f'(C)/(xM_{Pl})$. We can then use the scaling of mass to eliminate $x$. Thus, we get a relation for BS $f$-mode solely as a function of mass and compactness.

\begin{equation}\label{eqn:f_m_c_relation}
   \boxed{f_{BS} = f'(C)M'(C)\left(\frac{M_{Pl}^2}{M_{BS}}\right) = 2.03\times10^2f'(C)M'(C)\left(\frac{M_{\odot}}{M_{BS}}\right) \text{kHz}}~.
\end{equation}

\begin{figure}
    \centering
    \includegraphics[width=0.31\linewidth]{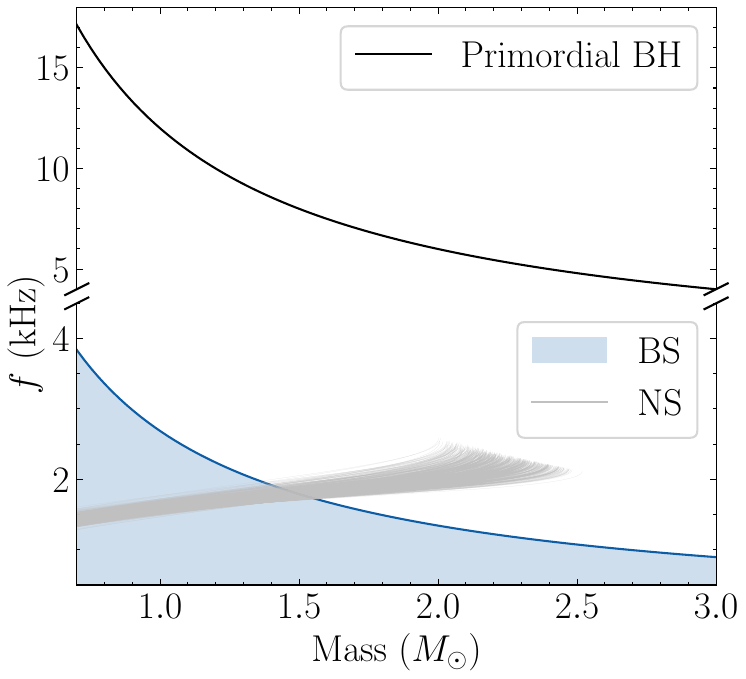}
    \label{fig:bh_qnm_ns}
    \includegraphics[width=0.31\linewidth]{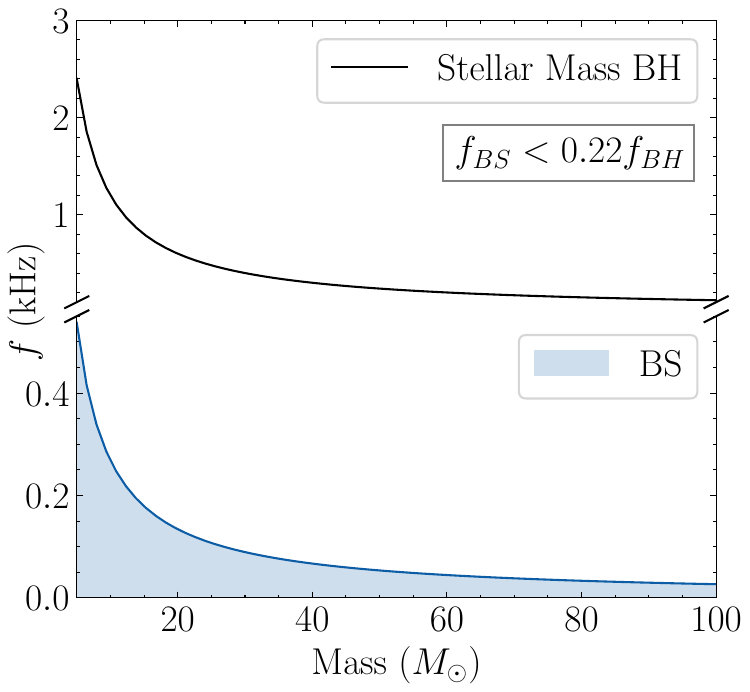}
    \label{fig:bh_qnm_bh}
    \includegraphics[width=0.34\linewidth, height=3.8cm]{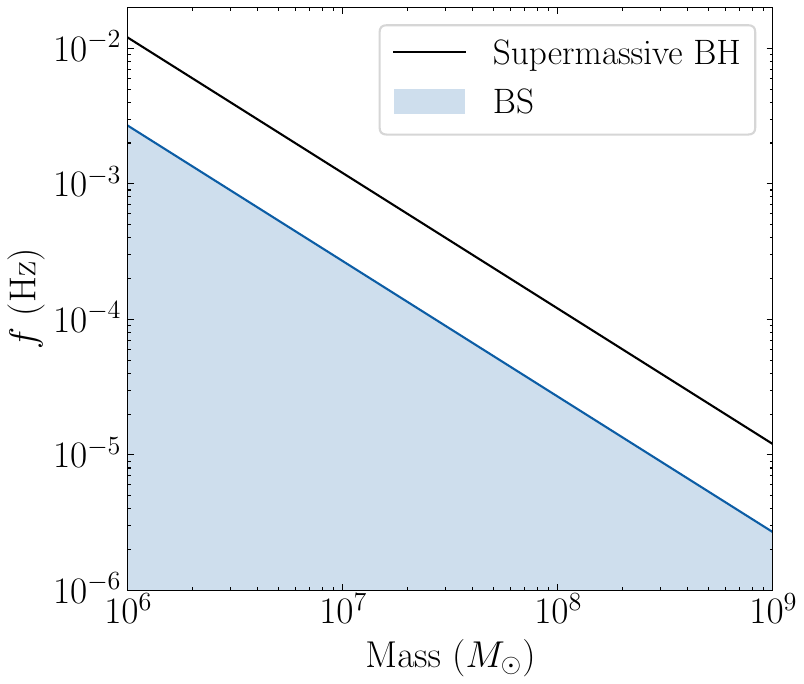}
    \label{fig:bh_qnm_smbh}
    \caption{$f$-mode frequency as a function of mass for (a) NS, (b) stellar-mass BH, and (c) SMBH mass range for massive BSs. The blue region is the part where BSs admit a solution. The blue curve corresponds to the case with maximum compactness. The black line is the corresponding BH QNM frequency~\citep{Kokkotas1999, Berti2009}. The BS frequencies satisfy $f_{BS}/f_{BH} < 0.22$ (see text for more details). The silver curves are indicative of the region spanned by NSs~\citep{Pradhan2022}.}
    \label{fig:comparison_bh_ns_qnm}
\end{figure}

This is a very useful relation as the frequency can be computed using just two physical parameters, i.e., $M$ and $C$. We can use the unique solutions obtained in Sec.~\ref{sec:f_modes} to evaluate $f'(C)$ and $M'(C)$ (see Fig.~\ref{fig:f_m_c}). 
    The blue regions in Fig.~\ref{fig:comparison_bh_ns_qnm} are obtained by varying the compactness for a given mass. The case of maximum compactness (dark blue curve in Fig.~\ref{fig:comparison_bh_ns_qnm}) is obtained by using $C=C_{max}$, arriving at the equation
\begin{align}\label{eqn:fmax_m_relation_2}
    \boxed{f_{BS,max} = 0.013 \left(\frac{M_{Pl}^2}{M_{BS}}\right)  = 2.6  \left(\frac{M_{\odot}}{M_{BS}}\right) \text{kHz}}~.
\end{align}
This only depends on the mass of the BS, as in the case of BHs.

The $f$-mode oscillations of NSs have been studied in great detail. Given the underlying uncertainty in NS EoS, we obtain a band in the $f$-$M$ space. NSs lie in the mass range of 1-2.5 $M_{\odot}$ with compactness in the range 0.1-0.3. We show this mass range in Fig.~\ref{fig:comparison_bh_ns_qnm} (a). The blue contour shows the region covered by BSs. The upper boundary corresponds to $C=C_{max}=0.16$. Silver curves show the $f-M$ region spanned by NSs due to the uncertainty in the microscopic equation of state (taken from~\cite{Pradhan2022}). We see a partial overlap with the NS band, but BS frequencies in the higher mass range are lower than those of NSs. Thus, GW observation of $f$-mode oscillations of a compact object in the NS mass range but outside the expected NS frequency band would be a signature for BS.

Astrophysical BHs are not expected in the mass range of NSs. However, PBHs, if they exist, may have mass in the solar and sub-solar mass range. On the other hand, stellar mass BHs fall in the range 5-100 $M_{\odot}$ and supermassive BHs (SMBH) in galactic centers have masses $10^6$-$10^9$ $M_{\odot}$. We show these two mass ranges in Figs.~\ref{fig:comparison_bh_ns_qnm} (b) and~\ref{fig:comparison_bh_ns_qnm} (c), respectively. BHs with masses larger than 100 $M_{\odot}$ are possible via mergers of lighter stellar mass BHs~\cite{IMBH2020}. We show the $n=0$, $l=2$ fundamental $f$-mode QNM frequency~\cite{Kokkotas1999} by the black curves in Fig.~\ref{fig:comparison_bh_ns_qnm}. This is given by $\omega M = 0.3737 - 0.0890i$~\cite{Kokkotas1999, Berti2009} in geometrical units ($G=c=1$). Reverting to natural units means
\begin{align}\label{eqn:bh_qnm_eqn}
    f_{BH} &= 0.059 \left(\frac{M_{Pl}^2}{M_{BH}}\right) = 1.2 \left(\frac{10M_{\odot}}{M_{BH}}\right) \text{kHz}~.
\end{align}
The BH frequency only depends on the mass of a non-rotating, chargeless BH (no hair). The BH frequency is also consistently higher than that of BS by a certain factor. This is expected from the high compactness of BH ($C=0.5$), which is higher by a factor of $3.125$ compared to the most compact BS. This factor is readily calculable using Eqs.~\ref{eqn:fmax_m_relation_2} and~\ref{eqn:bh_qnm_eqn}. We define a ratio ($\mathcal{R_{BS/BH}}$) as the ratio of the fundamental QNM frequency of the most compact BS to that of BH of the same mass. This ratio is a constant, independent of mass, and is given by
\begin{equation}
    \boxed{\mathcal{R_{BS/BH}} = \frac{f_{BS, max}}{f_{BH}} = 0.22}
\end{equation}
The BH frequency is higher than the most compact BS configuration of the same mass throughout by a factor of $0.22^{-1}=4.5$. We thus conclude that BS $f$-modes are always lower than BH frequencies of the same mass by a factor of at least 4.5, i.e., $f_{BS}/f_{BH} < 0.22$. If we observe $f$-modes from dark compact objects with a frequency consistently less than $\mathcal{R_{BS/BH}}$ times the expected BH frequency, it would be a smoking gun for massive BSs, which is the main takeaway of this analysis. Note that a similar analysis for other potentials ($V(\phi)$), as well as for another type of BSs, e.g., Oscillatons and Proca Stars, is necessary to distinguish between different models for stars made of Bosonic DM.  However, this still acts as a clear signature for distinguishing BHs from BSs. In general, for a BS of arbitrary compactness, the ratio is given by 
\begin{equation}
    \frac{f_{BS}}{f_{BH}} = 16.8f'(C)M'(C)~.
\end{equation} 

We make another comparison with BHs. A relation similar to Eq.~\ref{eqn:f_m_c_relation} exists for the innermost stable circular orbit (ISCO) frequency ($f_{ISCO}$)~\cite{Guidice2016}
\begin{equation}
    f^{ISCO} = \frac{1}{2\pi}\left(\frac{C}{3}\right)^{3/2}\left(\frac{M_{Pl}^2}{M_{BH}}\right)~.
\end{equation}
Here, $M$ is the component mass in an equal mass binary. The ISCO frequency marks the end of the inspiral phase during the binary merger. For $C=0.5$, we get the ISCO frequency for BH. We compare the $f$-mode frequency with the ISCO, as it marks the importance of dynamical tides during pre-merger in the case of a BS binary system. For BHs, the ratio is 
\begin{equation}
    \frac{f_{BH}}{f_{BH}^{ISCO}} = 0.3737\times6^{3/2} = 5.5~.
\end{equation}
Note that this number is not as important for BHs in the context of dynamical tides as it will be for BSs, since there are no BH tidal deformations. For BSs, we get
\begin{equation}
    \frac{f_{BS}}{f_{BS}^{ISCO}} = 2\pi f'(C)M'(C)\left(\frac{3}{C}\right)^{3/2}~.
\end{equation}
In the case of BSs, the ratio depends on compactness. The maximum value obtained for the most compact case is 6.6, higher than that for BH. The reason is that although the QNM frequency is higher for BH, the ISCO frequency is much lower in the case of BSs, which are much less compact. 

\section{Summary} \label{sec:summary}
We have proposed new scaling relations connecting global characteristics of fundamental modes in Boson stars (BS) with microscopic dark matter (DM) parameters. We considered a class of dark stars called Boson stars, modeled by a complex scalar field $\phi$ that interacts gravitationally and has self-interaction given by the potential $V(\phi)$. We focussed on the simple case of quartic potential given by $V(\phi) =\frac{1}{2}m^2\phi^2 + \frac{1}{4}\lambda \phi^4$. The strong-interaction limit ($\Lambda = \lambda M_{Pl}^2/ 4\pi m^2/ \gg 1$) leads to the formation of a ``massive star" whose mass scales as $\propto \sqrt{\lambda}M_{Pl}^3/m^2$, resembling massive fermionic stars. Such stars are compact objects and can act as BH mimickers, forming a fraction of the DM in the universe.
\\

The main highlights of this work:
\begin{itemize}
    \item We evaluated the $f$-mode characteristics for massive BSs within a full general-relativistic setup. We derived the $f$-mode frequency and damping time scaling relations and demonstrated for the first time that the scaled dimensionless $f$-mode characteristics ($f'$ and $\tau'$) are independent of the model parameters. 
    \item We report the relations $f'-M'$, $f'-C'$, and $\tau'-M'$, $\tau'-C'$, which are unique curves for massive BSs. The scaling relations $f=f'/(xM_{Pl})$ and $\tau = \tau'(xM_{Pl})$ can be used to evaluate the $f$-mode characteristics for any parameter values in the available parameter space. 
    \item $f'$ is an increasing function of $M'/C'$ whereas $\tau'$ is a decreasing one. Thus, the configuration corresponding to the maximum $M'=0.062$ or $C'=0.16$  was used to derive the maximum value of $f'$ given by $f'_{max}=0.21$ and a minimum value of $\tau'$ given by $\tau'_{min}=1900$. This also results in a minimum number of BS $f$-mode oscillations given by $N_{min}=406$.
    \item The derivation of the scaling relations allows the computation of BS QNMs for any parameters and enables a systematic comparison of $f$-mode parameters with other compact objects. We compared the BS $f$-mode frequencies with those of NSs and BH $l=2$ QNMs. We derived the expression for BS $f$-mode frequency that depends solely on the mass and compactness. 
    \item For the maximum compactness configuration, the frequency is the highest and only depends on the mass, as in the case of BHs. For such configurations, the BS $f$-mode frequency was shown to be lower than the corresponding BH frequency by a constant factor given by $\mathcal{R_{BS/BH}} = 0.22$. Thus, observing BH mimickers with $f$-mode frequency consistently lower than 0.22 times that of the expected BH frequency would be a clear signature of massive BS. 
    \item We also compared the BS $f$-mode frequency with the corresponding ISCO frequency and found that the ratio of $f$-mode frequency to that of ISCO is larger for BSs compared to BHs, given their low compactness.
\end{itemize}

\cite{Flores2019} is the first work to explore non-radial fundamental modes for massive BSs in the strong-coupling limit, followed by~\cite{Celato2025}. These works explored BS configurations for select model parameters. The scalings with $x$ introduced in this work can be used to compute $f$-mode characteristics for any model parameters using a single solution, establishing that there is no need to study $f$-modes separately for different parameter values.
A list of observational signatures for BSs and a comparison of $f_{ISCO}$ between ECOs and BHs was provided in~\cite{Guidice2016}. This work extends the list to include the fundamental QNM frequencies as a tool to distinguish BSs from NSs and BHs. 

The results presented here can be used for searches of massive BS in the strong-interaction limit where the effects of $f$-mode oscillations are important. These would be relevant in the future when $f$-mode oscillations become detectable. With upcoming GW detectors across various frequency bands like LISA (0.1 mHz-1 Hz), LIGO (50 Hz-1000 Hz), and NEMO (2 kHz-4 kHz), a wide range of BS masses could be probed. As the $f$-mode frequency depends on compactness and mass, we get that for $C\gtrsim0.01$, BSs roughly falling in the mass range of $10^2-10^7M_{\odot}$ would be observable by LISA, $0.1-10M_{\odot}$ by
LIGO and $0.01-1M_{\odot}$ by NEMO. The next-generation detectors CE and ET roughly encompass the range probed by LIGO and NEMO.
Although tidal deformability remains the dominant distinguishing parameter in a binary inspiral~\cite{Sennett2017}, as it was shown that incorporating higher order effects like the spin-induced quadrupole moment improves the distinguishability~\cite{Krishnendu2017, Krishnendu2019} and the constraining power of BS microscopic parameters~\cite{Pacilio2020, Vaglio2023}, including the effect of $f$-mode oscillations has the potential to improve it further. Distinguishing from NSs in binary remains a challenge, as BSs follow the same $f$-Love universal relations. However, the results are particularly useful as a distinguishability tool in the context of post-merger GW spectroscopy and in future detections of GW bursts from oscillations of isolated systems. Further analyses on the detectability of these $f$-modes are essential. A detailed study of their detectability from isolated GW bursts has been carried out in a separate work~\cite{shirke_BS_fmodes_long}.

In this work, we have compared BS $f$-modes with those of NSs and BHs. This can be further extended to include NSs with exotic matter such as hybrid stars, strange stars, DM admixed NSs, etc. Here we considered scalar DM with $\phi^4$ potential resulting in $C_{max} \approx 0.16$. Higher $C_{max}$ can be achieved with $\phi^n$ potential for $n>4$~\cite{Pitz2023} possibly resulting in a higher $f'_{max}$ and a higher $\mathcal{R}$. This can lead to configurations filling the gap between NSs and BHs (see Fig.~\ref{fig:comparison_bh_ns_qnm}), making it distinguishable from NSs as well. A complete analysis of ECOs would also require consideration of other scalar DM potentials, vector DM, and other ECOs such as fermion stars, gravastars, wormholes, fuzzballs, etc.~\cite{CardosoPani2019}.

\section*{Acknowledgements}

L.S. and J.S.B. acknowledge support by the
Deutsche Forschungsgemeinschaft (DFG, German Research Foundation) through the CRC-TR 211 `Strong interaction matter under extreme conditions' – project
no. 315477589 – TRR 211.

\appendix

\section{Scaling of $f$-mode equations: Cowling Formalism}\label{sec:appendix_fmode_cowling}
We will now see how the $f$-mode equations and the $f$-mode parameters get modified when scaling and scaled EoS are applied. We first look at the equations in the Cowling approximation, where metric perturbations are neglected, assuming a weak gravitational field. We do not use the Cowling formalism in this work as we solve the equations in full general relativity. However, we demonstrate that the scaling holds for solutions in Cowling approximation as well, for the sake of completeness and the benefit of any future works. 

The spacetime metric for a spherically symmetric background is given by
 \begin{equation}
     ds^2=-e^{2\Phi (r)}dt^2+e^{2\Lambda (r)}dr^2+r^2 d\theta^2+r^2\sin^2{\theta} d\phi^2 \label{eqn:metric}
 \end{equation}

Here, 
\begin{align}
    \frac{d \Phi(r)}{dr}&=\frac{-1}{\epsilon(r)+p(r)}\frac{dp}{dr} \\
    \frac{d \Phi'(r')}{dr'}&=\frac{-1}{\epsilon'(r')+p'(r')}\frac{dp'}{dr'}
\end{align}

Thus, $\Phi' = \Phi$. Similarly, $\Lambda'=\Lambda$. This metric function $\Lambda(r)$ is different from the dimensionless tidal deformability. The differential equations to solve for the $f$-mode frequencies are given by~\cite{Sotani2011,Flores2014, Sandoval2018, Pradhan2021}:
\begin{align}
    \frac{d W(r)}{dr}&=\frac{d \epsilon}{dp}\left[\omega^2r^2e^{\Lambda (r)-2\Phi (r)}V (r)+\frac{d \Phi(r)}{dr} W (r)\right] \nonumber \\
    &- l(l+1)e^{\Lambda (r)}V (r) \\
    \frac{d V(r)}{dr} &= 2\frac{d\Phi (r)}{dr} V (r)-\frac{1}{r^2}e^{\Lambda (r)}W (r) 
    \label{eqn:perteq_1}
\end{align}
The functions $V (r)$ and $W (r)$ are associated with the Lagrange displacement vector of the perturbed fluid. Now we write these in terms of our new coordinates, $r = r'xM_{Pl}$. We see the $\omega$ (the $f$-mode frequency) appears as the product ($\omega r$). So we scale $\omega$ as inverse of $r$, $\omega = \omega'/(xM_{Pl})$ such that $\omega r = \omega' r$. The equation becomes:
\begin{align}
    \frac{d W(r)}{dr'xM_{Pl}}&=\frac{d \epsilon'}{dp'}\left[\omega'^2r'^2e^{\Lambda' (r')-2\Phi' (r')}V (r)+\frac{d \Phi'(r')}{dr'xM_{Pl}} W (r)\right] - l(l+1)e^{\Lambda' (r)}V (r)  \\
     \frac{d V(r)}{dr'xM_{Pl}} &= 2\frac{d\Phi' (r')}{dr'xM_{Pl}} V (r)-\frac{1}{r'^2x^2M_{Pl}^2}e^{\Lambda' (r')}W (r) 
    \label{eqn:perteq_2}
\end{align}

The solutions at the center of the star are given by the following ansatz
\begin{align}
    W (r)=Ar^{l+1},& \> V (r)=-\frac{A}{l} r^l~, \\
    \implies W' (r')=Ar'^{l+1},& \> V' (r')=-\frac{A}{l} r'^l~,
\end{align}
where $A$ is the amplitude of oscillations. We see that $W$ scales as $r^{l+1}$ and $V$ as $r^l$. We use, $W(r)=W'(r')(xM_{Pl})^{l+1}$ and $V(r)=V'(r')(xM_{Pl})^l$. Adopting this scaling, we find the primed equations remain unchanged and are free of $x$. Thus, the solutions do not depend on model parameters. The desired eigenfrequency is obtained by imposing the boundary condition
\begin{align} 
    \omega^2e^{\Lambda (R)-2\Phi (R)}V (R)&+\frac{1}{R^2}\frac{d\Phi (r)}{dr}\Big|_{r=R}W (R)=0~, \\
    \implies \omega'^2e^{\Lambda' (R')-2\Phi' (R')}V' (R')&+\frac{1}{R'^2}\frac{d\Phi' (r)'}{dr'}\Big|_{r=R}W' (R')=0~,
\label{eqn:bc}
\end{align}
which comes from vanishing perturbed Lagrangian pressure at the surface.  Thus, we can solve the eigenfrequencies $\omega'$ in primed coordinates (EoS independent) for different configurations $M'$ or $R'$ and use the $\omega = \omega'/(xM_{Pl})$ to get the $f$-mode frequencies for BSs with arbitrary $(\lambda, m)$.

\section{Scaling of $f$-mode equations: Full General Relativity}\label{sec:appendix_fmode_gr}
Here, we present the equations used to solve the complex QNM frequencies using full general relativity, where metric perturbations are coupled to matter. Then we provide the scaling of various quantities that make the equations independent of the model parameters.
\subsection{Perturbations Inside the Star}
The perturbed metric ($ds^2_p$) can be written as~\cite{Thorne},
\begin{eqnarray}
    ds^2_p=ds^2+h_{\mu \nu} dx^{\mu}dx^{\nu}~.
    \label{eqn:perturbedmetric}
\end{eqnarray}

For the even-parity (polar) perturbations where the GW and matter perturbations are coupled, the $h_{\mu \nu}$ can be expressed as~\cite{Sotani2001,Thorne},
\begin{eqnarray}
    h_{\mu \nu}=
   \begin{pmatrix}
r^lHe^{2\Phi} & i\omega r^{l+1} H_1 &0&0\\
i\omega r^{l+1} H_1 & r^l H e^{2\Lambda} &0 &0\\
0 &0 & r^{l+2}K& 0\\
0 & 0& 0 &  r^{l+2}K sin^2{\theta}
\end{pmatrix} Y^l_m e^{i\omega t} ~, \nonumber\\
 \text{ }
 \label{eqn:metricfunctions}
\end{eqnarray}
where $Y_m^l$ are spherical harmonics. $H,\ H_1, \ K$ are perturbed metric functions and are functions of the radial coordinate $r$.

The  Lagrangian displacement vector  $\textbf{$\zeta$}=(\zeta^r,\zeta^{\theta},\zeta^{\phi})$  for the polar perturbations of the fluid is given by,

\begin{equation}
    \zeta^{r}=\frac{r^l}{r}e^{-\Lambda} W(r)  Y^l_m e^{i\omega t}~;\quad
    \zeta^{\theta}=\frac{-r^l}{r^2} V(r)  \frac{\partial Y^l_m}{\partial \theta} e^{i\omega t}~; \quad
    \zeta^{\phi}=\frac{-r^l}{r^2 sin^2\theta} V(r)  \frac{\partial Y^l_m}{\partial \phi} e^{i\omega t}~,
    \label{eqn:pertfluid}
\end{equation}
where $W, V$ are amplitudes of the radial and transverse fluid perturbations. The equations governing these perturbation functions and the metric perturbations inside the star are given by~\cite{Sotani2001},
\begin{eqnarray}
    \frac{d H_1}{dr}&=&\frac{-1}{r}\left[l+1+\frac{2m}{r}e^{2\Lambda}+4\pi r^2e^{2\Lambda} G\left( p-\epsilon \right)\right] H_1 \nonumber \\ 
    &&+ \frac{1}{r}e^{2\Lambda}\left[H+K+16\pi G\left(p+\epsilon\right)V\right] \label{eqn:dh1} \ , \\
    \frac{d K}{dr}&=&\frac{l(l+1)}{2r}H_1+\frac{1}{r}H-\left(\frac{l+1}{r}-\frac{d\Phi}{dr}\right)K+\frac{8\pi}{r}G\left(p+\epsilon\right)e^{\Lambda} W \ ,  \label{eqn:dk} \\
    \frac{d W}{dr}&=&re^{\Lambda}\left[\frac{1}{\gamma Gp}e^{-\Phi}X-\frac{l(l+1)}{r^2}V-\frac{1}{2}H-K\right] -\frac{l+1}{r}W   \label{eqn:dw} \ , \\
    \frac{d X}{dr}&=& \frac{-l}{r}X+G(p+\epsilon)e^{\Phi}\Bigg[\frac{1}{2}\left(\frac{d\Phi}{dr}-\frac{1}{r}\right)H -\frac{1}{2}\left( \omega^2re^{-2\Phi}+\frac{l(l+1)}{2r}\right)H_1 \nonumber \\
    &+&\left(\frac{1}{2r}-\frac{3}{2}\frac{d\Phi}{dr}\right)K - \frac{1}{r}\left[ \omega^2\frac{e^{\Lambda}}{e^{2\Phi}}+4\pi G(p+\epsilon ) e^{\Lambda}-r^2\frac{d}{dr}\left( \frac{e^{-\Lambda}}{r^2}\frac{d \Phi}{dr}\right)\right]W \nonumber \\
    &-&\frac{l(l+1)}{r^2}\frac{d\Phi}{dr}V\Bigg]  \label{eqn:dx}\ ,
\end{eqnarray}
\begin{eqnarray}
   &&\left[1-\frac{3m}{r}-\frac{l(l+1)}{2}-4\pi r^2Gp\right]H-8\pi r^2 e^{-\Phi}X \nonumber \\ &+&r^2e^{-2\Lambda}\left[\omega^2e^{-2\Phi}-\frac{l(l+1)}{2r}\frac{d\Phi}{dr}\right]H_1 \nonumber \\
   &-& \left[ 1+ \omega^2r^2e^{-2\Phi} -\frac{l(l+1)}{2}-(r-3m-4\pi r^3Gp)\frac{d\Phi}{dr}\right]K =0  \label{eqn:h}\\
  && e^{2\Phi}\left[ e^{-\Phi}X+\frac{e^{-\Lambda}}{r}G\frac{dp}{dr} W+\frac{(p+\epsilon)}{2}H\right] -\omega^2 G(p+\epsilon) V=0 ~, \label{eqn:v}
\end{eqnarray}

where $X$ is introduced as~\cite{Detweiler83,Sotani2001}
\begin{eqnarray}
    X&=&\omega^2G(p+\epsilon ) e^{-\Phi} V-\frac{We^{\Phi-\Lambda}}{r}G\frac{dp}{dr}-\frac{1}{2} G(p+\epsilon ) e^{\Phi}H\,, \nonumber \\
    \label{eqn:x}
\end{eqnarray}
$m=m(r)$ is the enclosed mass of the star and $\gamma$ is the adiabatic index defined as
\begin{equation}
    \gamma=\frac{(p+\epsilon)}{p}\left(\frac{\partial p}{\partial \epsilon }\right)\bigg|_{ad} ~.
     \label{eqn:gamma}
\end{equation}

While solving these equations, we have to impose boundary conditions, i.e., the perturbation functions are finite throughout the interior of the star (particularly at the center, i.e., at $r=0$), and the perturbed pressure ($\Delta p$) vanishes at the surface.  Function values at the center of the star can be found using the Taylor series expansion method described in Appendix B of~\cite{Detweiler83} (see also  Appendix A of~\cite{Sotani2001}). The vanishing perturbed pressure at the stellar surface is equivalent to the condition $X(R)=0$ (as $\Delta p=-r^l e^{-\Phi}X$). We follow the procedure described in~\cite{Detweiler83} to find the unique solution for a given value of $l$ and $\omega$ satisfying all the boundary conditions inside the star.

\subsection{Perturbations outside the star and complex eigenfrequencies}
The perturbations outside the star are described by the Zerilli  equation~\cite{Zerilli}.
\begin{equation}
    \frac{d^2Z}{dr_*^2}+\omega^2 Z=V_Z Z
    \label{eqn:zerilli}
\end{equation}
 where $r_*=r+2m \log \left({\frac{r}{2m}-1}\right)$ is the tortoise co-ordinate and $V_Z$ is defined as ~\cite{Zerilli},
 \begin{eqnarray}
     V_Z&=&\frac{2(r-2m)}{r^4 (nr+3m)^2}\Big[n^2(n+1)r^3 +3n^2mr^3+9nm^2r+9m^3\Big] ~,
 \end{eqnarray}
where $n=\frac{1}{2} (l+2)(l-1)$. Asymptotically the wave solution to \eqref{eqn:zerilli} can be expressed as \eqref{eqn:zerillisolution},
\begin{eqnarray}
    Z=A(\omega)Z_{in}&+&B(\omega) Z_{out}\,, \label{eqn:zerillisolution}\\
    Z_{out}=e^{-i\omega r^*} \sum_{j=0}^{j=\infty}\alpha_j r^{-j}&,& Z_{in}=e^{i\omega r^*} \sum_{j=0}^{j=\infty}\bar{\alpha}_j r^{-j} ~.\nonumber
    \end{eqnarray}
 Keeping  terms up to $j=2$ one finds,
 \begin{equation}
     \alpha_1=-\frac{i}{\omega}(n+1)\alpha_0,~; \quad
     \alpha_2=\frac{-1}{2\omega^2}\left[n(n+1)-i3M\omega\left(1+\frac{2}{n}\right)\right]\alpha_0
 \end{equation}

For initial boundary values of Zerilli functions, we use the method described in ~\cite{Fackerell,Detweiler85,Sotani2001}. Setting $m=M$ and perturbed fluid variables to 0 (i.e., $W=V=0$) outside the star, connection between the metric functions \eqref{eqn:metricfunctions} with Zerilli function ($Z$ in Eq.\eqref{eqn:zerilli}) can be written as,
\begin{eqnarray}
    \begin{pmatrix}
    r^lK\\
    r^{l+1}H_1
    \end{pmatrix}
    &=&
    \begin{pmatrix}
    \frac{ n(n+1)r^2+3nMr+6M^2}{r^2(nr+3M)} & 1\\
    \frac{nr^2-3nMr-3M^2}{(r-2M)(nr+3M)} & \frac{r^2}{r-2M}
    \end{pmatrix}
    \begin{pmatrix}
    Z \\
    \frac{dZ}{dr_*}
    \end{pmatrix}
    \label{eqn:zerillconnection}
\end{eqnarray}

The initial boundary values of Zerilli functions are fixed using \eqref{eqn:zerillconnection}. The Zerilli equation \eqref{eqn:zerilli} is then integrated numerically to infinity, and the complex coefficients $A(\omega),\ B(\omega)$ are obtained by matching the analytic expressions for $Z$ and $\frac{dZ}{dr_*}$ with the numerically obtained value of $Z$ and $\frac{dZ}{dr_*}$. The natural frequencies that are not driven by incoming gravitational radiation represent the quasi-normal mode frequencies. Mathematically, we find the complex roots of  $A(\omega)=0$, representing the complex eigenfrequencies of  QNMs.

\subsection{Scaling}
As the radius of the massive BS in the strong interaction limit scales as $R=R'(xM_{Pl})$, we scale the radial coordinate as $r=r'(xM_{Pl})$. The distance between two points also then becomes $ds=ds'(xM_{Pl})$. Thus, the whole spatial scale is scaled up.
$h_{\mu \nu}$, which is dimensionless, then remain unaltered under $r=r'(xM_{Pl})$, $\theta'=\theta$ and $\phi'=\phi$. This is achieved if the quantities $H$ and $K$ scale as $H = H'/(xM_{Pl})^l$ and $K = K'/(xM_{Pl})^l$, $h_{\mu \nu} = h_{\mu \nu}'$. Thus $H$ and $K$ scale as $1/r^l$.
Using these and the perturbation equations, we arrive at the following transformations:\\

\begin{center}
\begin{tabular}{|c|c|c|}
\hline
\multicolumn{3}{|c|}{Scalings} \\
\hline
$ds=ds'(xM_{Pl})$ & $H = H'/(xM_{Pl})^l$ & $m=m'(xM_{Pl})$\\
\hline
$h_{\mu \nu} = h_{\mu \nu}'$ & $K = K'/(xM_{Pl})^l$ & $p=p'/x^2$\\
\hline
$r=r'(xM_{Pl})$ & $H_1 = H_1'/(xM_{Pl})^l$ & $\epsilon=\epsilon'/x^2$\\
\hline
$t=t'(xM_{Pl})$    & $Y^{l}_m=Y^{l \prime}_m$  & $\gamma=\gamma'$\\
\hline
$\theta'=\theta$   & $W=W'/(xM_{Pl})^{l-2}$ & $\omega=\omega'/(xM_{Pl})$\\
\hline
$\phi'=\phi$ & $V=V'/(xM_{Pl})^{l-2}$ & $X=X'/(xM_{Pl})^{l+2}$\\
\hline
$\Lambda=\Lambda'$ & $\zeta^r = \zeta^{r \prime}(xM_{Pl})$ & $Z=Z'(xM_{Pl})$  \\
\hline
$\Phi=\Phi'$ & $\zeta^\theta=\zeta^{\theta \prime}$ & $V_{Z} = V_{Z}'$\\
\hline
- &  $\zeta^\phi=\zeta^{\phi \prime}$ & - \\
\hline
\end{tabular}
\label{table:ta}
\end{center} 

Applying these transformations and using $G=1/M_{Pl}^2$ in all the above equations, we find that they can be written completely in terms of the primed quantities, and all the factors of $x$ cancel out. Initially, the $x$ dependence was via the equation of state $p$ and $\epsilon$. Now $p'$ and $\epsilon'$ do not depend on $x$. Thus, the equations are completely free of the model parameters. For each central density $\rho'_c$ or, equivalently, mass $M'$, we get a unique solution for $\omega'$. The scaling relation $\omega=\omega'/(xM_{Pl})$ can then be used to obtain the $f$-mode frequency and damping time for any BS with DM model parameters $(\lambda,m)$, i.e., $\omega(x, M) = \omega'(M'=M/(xM_{Pl}^3))$. $\omega \propto f+i/\tau$. Thus we can write $f=f'/(xM_{Pl})$ and $\tau =\tau'(xM_{Pl})$.

It is well-known that the quadrupolar fundamental non-radial frequency scales linearly with the square root of the average density of star~\cite{AnderssonKokkotas1996,AnderssonKokkotas1998} ($f \sim \sqrt{GM/R^3}$). From the scaling of $M$ and $R$ mentioned above, we can say that the scaling of $f$ mentioned above is consistent with this.
Thus, $f$ scales as the inverse of $r$.

Furthermore, the damping time should follow (chapter 36.2 of~\cite{MTW2017}) $\tau \sim R^4/(GM)^3$, which is also consistent with the derived scaling for $\tau$.
The damping time is defined to be the inverse of the imaginary part of the complex eigenfrequency, so it should scale as the inverse of $f$. Moreover, the mass-scaled damping time ($GM/\tau$) shows universality with stellar compactness $C=M/R$~\cite{Tsui2005}. Hence, the initial guess would be that  $\tau$ can be scaled as $r$.

 \bibliographystyle{elsarticle-num} 
 \bibliography{refs}

\end{document}